\documentstyle[12pt]{article}
\newcommand{\beq}{\begin{equation}}
\newcommand{\eeq}{\end{equation}}
\newcommand{\bea}{\begin{eqnarray}}
\newcommand{\eea}{\end{eqnarray}}

\begin{document}
\setcounter{page}{0}
\topmargin 0pt
\oddsidemargin 5mm
\renewcommand{\thefootnote}{\fnsymbol{footnote}}
\newpage
\setcounter{page}{0}
\begin{titlepage}
\begin{flushright}
QMW 94-9
\end{flushright}
\begin{flushright}
hep-th/9405016
\end{flushright}
\vspace{0.5cm}
\begin{center}
{\large {\bf Perturbed Gauged WZNW Models}} \\
\vspace{1.8cm}
\vspace{0.5cm}
{\large Oleg A. Soloviev
\footnote{e-mail: soloviev@V1.PH.QMW.ac.uk}\footnote{Work supported by
S.E.R.C. and in part by a contract from the European Commission Human Capital
and Mobility Programme.}}\\
\vspace{0.5cm}
{\em Physics Department, Queen Mary and Westfield College, \\
Mile End Road, London E1 4NS, United Kingdom}\\
\vspace{0.5cm}
\renewcommand{\thefootnote}{\arabic{footnote}}
\setcounter{footnote}{0}
\begin{abstract}
{We discuss a new type of unitary perturbations around conformal theories
inspired by the $\sigma$-model perturbation of the nonunitary WZNW model.
We show that the nonunitary level $k$ WZNW model perturbed by its sigma model
term goes to the unitary level $-k$ WZNW model. When plugged into the gauged
WZNW model the given perturbation results in the perturbed gauged WZNW model
which no longer describes a coset construction. We
consider the BRST invariant generalization of the sigma model perturbation
around the gauged WZNW model. In this way we obtain perturbed coset
constructions.
In the case of the $SU_{m-2}(2)\times SU_1(2)/SU_{m-1}(2)$ coset, the BRST
invariant sigma model perturbation is identical to Zamolodchikov's
$\Phi_{(3,1)}$ perturbation of the minimal conformal series. The existence of
general geometry flows is clarified.}
\end{abstract}
\vspace{0.5cm}
\centerline{April 1994}
 \end{center}
\end{titlepage}
\newpage
%******************************************************************
\section{Introduction}

There has recently been much interest attracted to gauged
Wess-Zumino-Novikov-Witten (GWZNW) models
\cite{Bardakci_Rabinovici}-\cite{Karabali_Schnitzer} which turn out to give
rise to exact (to all orders in $\alpha'$) nontrivial string solutions
\cite{Bardakci_Crescimanno}-\cite{Klimcik-1}. Such an outstanding situation
occurs because of the fact that GWZNW models provide a proper Lagrangian
formulation to algebraic coset constructions
\cite{Bardakci_Halpern}-\cite{Goddard} which are exactly solvable conformal
field theories. On the other hand there are examples of many other exact
conformal field theories which are distinct from cosets.
One possible route away
from coset constructions may be through some relevant perturbations of
gauged WZNW models.
Therefore, there may exist some room for possibly new exact string backgrounds
by perturbations of GWZNW models.

All string solutions are believed to be points in the coordinate space of
string field theory \cite{Zwiebach}, which is supposed to be
independent of any particular choice of coordinates in analogy with general
relativity. Such a string field theory formulation, though still being
developed, apparently necessitates this coordinate space to be a connected
multitude. Obviously conformal models alone cannot form any connected multitude
of string solutions. Therefore, the whole configuration space of string field
theory is going to be much more complex than the space of conformal models.

Some interesting properties of this configuration space have been analyzed in
\cite{Kutasov}-\cite{Cvetic}. The main tools which appear to be very
effective in elaborating the features of this highly nontrivial object are the
methods of perturbed conformal field theory
\cite{Zamolodchikov-1}-\cite{Cardy_Ludwig}. From the obtained results, the
coordinate
space of string field theory comes along to have the structure of a group-like
manifold whose local properties are determined by the algebra of all integrable
relevant deformations around conformal (Lagrangian) theories.
Unfortunately the knowledge of the
algebra of all relevant integrable perturbations around conformal Lagrangian
(unitary) models is still incomplete.
Accomplishing these tasks may turn out to need considerable effort.

In these notes we would like to consider some special relevant unitary
deformations around (gauged) WZNW models some of which have a clear space-time
interpretation. Some of the results were presented in
\cite{Soloviev-1}-\cite{Soloviev-3}. In the present paper we will describe
those results in more details and will discuss a new type of relevant
perturbations of the gauged WZNW model.

The perturbations we are going to exploit are generated by the sigma model term
of the nonunitary WZNW model \cite{Soloviev-1}. In spite of the nonunitarity of
the underlying WZNW model, the perturbing operator provides one with a unitary
Virasoro representation, as will be shown in section 2. The nonunitary WZNW
model perturbed by this operator will be demonstrated to go to the unitary WZNW
model. The theory, in which the nonunitary WZNW model emerges as an intrinsic
element, is the GWZNW model. In section 3 we will apply our perturbation to the
GWZNW model. In section 4 we will define a BRST invariant perturbing operator
which may act on the coset projection of the GWZNW model. In this way we will
obtain perturbed coset
constructions. We will exhibit that in the case of the $SU_{m-2}(2)\times
SU_1(2)/SU_{m-1}(2)$ GWZNW model, our BRST invariant perturbation reproduces
the
known deformation of the minimal conformal models by the $\Phi_{(3,1)}$
operator. In section 5 we will summarize our results and comment on them.

\section{The sigma model perturbation of the nonunitary WZNW model}

Our starting point is the level $k$ WZNW model \cite{Novikov}-\cite{Knizhnik}
described by the action
\begin{equation}
S_{WZNW}(g;k)={-k\over4\pi}\int
[\mbox{Tr}|g^{-1}\mbox{d}g|^2~+~{i\over3}\mbox{d}^{-1}\mbox{Tr}(g^{-1}\mbox{d}
g)^3],\end{equation}
where $g$ is the matrix field taking its values on the Lie group $G$. The
theory possesses the affine symmetry $\hat G\times\hat G$ which entails an
infinite number of conserved currents \cite{Witten},\cite{Knizhnik}. The latter
can be derived from the basic currents $J$ and $\bar J$,
\begin{eqnarray}
J&=&J^{a}t^{a}=-{k\over2}g^{-1}\partial g,\nonumber\\ & & \\
\bar J&=&\bar J^{a}t^{a}=-{k\over2}\bar\partial g\cdot g^{-1},\nonumber
\end{eqnarray}
satisfying the equations of motion
\begin{equation}
\bar\partial J=0,~~~~~~~~~\partial\bar J=0.\end{equation}
In eqs. (2.2) $t^{a}$ are the generators of the Lie algebra ${\cal G}$
associated with the Lie group $G$,
\begin{equation}
[t^{a},t^{b}]=f^{abc}t^{c},\end{equation}
with $f^{abc}$ the structure constants.

The important observation which has been made in \cite{Knizhnik} is that the
spectrum of the WZNW model contains states which correspond to the
primary fields of the
underlying affine symmetry. By definition, $\phi_i$ is an affine primary field,
if it has the following operator product expansion (OPE) with the affine
current $J$ \cite{Knizhnik}
\begin{equation}
J^a(w)\phi_i(z,\bar z)={t^a_i\over w-z}\phi_i(z,\bar z)~+~reg.,\end{equation}
where the matrices $t^a_i$ correspond to the left representation of
$\phi_i(z,\bar z)$. In the WZNW model, any affine primary field is Virasoro
primary and its conformal dimensions are given by \cite{Knizhnik}
\begin{equation}
\Delta_i={c_i\over c_V+k},~~~~~~~\bar\Delta_i={\bar c_i\over
c_V+k},\end{equation}
where $c_i=t^a_it^a_i,~\bar c_i=\bar t^a_i\bar t^a_i$ and $c_V$ is defined
according to
\begin{equation}
f^{acd}f^{bcd}=c_V\delta^{ab}.\end{equation}

The point to be made is that there are Virasoro primary fields in the spectrum
of the WZNW model which are not affine primary fields, but their
descendants. Our examples of such fields will be restricted to one particular
composite state.

Let us consider the following composite field
\begin{equation}
O=:J^a\bar J^{\bar a}\phi^{a\bar a}:\end{equation}
which is defined as a normal ordered product of the affine currents $J^a,~\bar
J^{\bar a}$ with the spin (1,1) affine-Virasoro primary field in the adjoint
representation of $G\times G$. The product of the three operators in eq. (2.8)
can be properly defined according to
\begin{equation}
O(z,\bar z)=\oint {\mbox{d}w\over2\pi i}~\oint {\mbox{d}\bar w\over2\pi i}
{J^a(w)\cdot\bar J^{\bar a}(\bar w)\cdot\phi^{a\bar a}(z,\bar z)\over|z-w|^2},
\end{equation}
where in the numerator the product is understood as an OPE. It is easy to see
that the given product does not contain singular terms.

{}From the definition it follows that the operator $O$ is an affine
descendant of the affine-Virasoro primary field $\phi$. Indeed, $O$ can be
presented in the form
\begin{equation}
O(0)=J^a_{-1}\bar J^{\bar a}_{-1}\phi^{a\bar a}(0),\end{equation}
where
\begin{equation}
J^a_{-1}=\oint {\mbox{d}w\over2\pi i}w^{-1}J^a(w),~~~~~~~~
\bar J^{\bar a}_{-1}=\oint {\mbox{d}\bar w\over2\pi i}\bar w^{-1}\bar J^{\bar
a}
(\bar w).
\end{equation}
At the same time, the operator $O$ continues to be a Virasoro primary operator.
Indeed, one can check that the state $O(0)|0\rangle$ is a highest weight
vector of the Virasoro algebra, with $|0\rangle$ the $SL(2,C)$ invariant
vacuum. That is,
\begin{eqnarray}
L_0O(0)|0\rangle&=&\Delta~O(0)|0\rangle,\nonumber\\ & & \\
L_{m>0}O(0)|0\rangle&=&0.\nonumber\end{eqnarray}
Here the generators $L_n$ are given by the contour integrals
\begin{equation}
L_n=\oint {\mbox{d}w\over2\pi i}w^{n+1}T(w),\end{equation}
where $T(w)$ is holomorphic component of the Sugawara stress tensor of the
conformal WZNW model,
\begin{equation}
T(z)={:J^a(z)J^a(z):\over k+c_V}.\end{equation}
In eqs. (2.12), $\Delta$ is the conformal dimension of the operator $O$. We
find
\begin{equation}
\Delta=\bar\Delta=1+{c_V\over k+c_V}.\end{equation}
Here $\bar\Delta$ is the conformal dimension of $O$ associated with
antiholomorphic conformal transformations.

Let us turn to the large $k$ limit. Then the operator $O$  becomes a
quasimarginal operator with anomalous dimensions
\begin{equation}
\Delta=\bar\Delta=1+{c_V\over k}+{\cal O}(k^{-2}).\end{equation}
When $k$ is negative, the given operator $O$ is relevant; whereas for
positive $k$ it is an
irrelevant operator. Relevant quasimarginal operators are of a great
interest because they can be used as perturbing operators around given
conformal theories.

The point to be made is that the WZNW model with negative $k$ is a nonunitary
theory, that is there are states with negative norms in the spectrum and the
Hamiltonian is unbounded below. At the
same time the theory is not completely meaningless. We will show that there are
states which have positive norms with respect to the $SL(2,C)$ vacuum. Thus,
there exists a subsector in which the theory can be properly defined by means
of a
certain truncation of the fusion algebra. Note that $|k|$ is thought of being
integer and, hence, such a truncation must exist.

In the large negative $k$ limit, the operator $O$
appears to be suitable for performing a perturbation around the nonunitary WZNW
model. The significant point to be made is that despite the nonunitarity of the
conformal model to be perturbed, the perturbation operator $O$ corresponds to a
unitary highest weight vector of the Virasoro algebra.
Indeed, as we have shown above, $O$ has positive conformal
dimensions. In turn, the Virasoro central charge of the nonunitary WZNW
model in the large negative level limit is bigger than one,
\begin{equation}
c_{WZNW}(k)={k\dim G\over k+c_V}=\dim G~+~{\cal O}(1/k)>1.\end{equation}
In fact, the operator $O$ lies in the unitary range of the Kac
determinant and, hence, it provides a unitary representation of the Virasoro
algebra\footnote{In fact, there are many other operators in the spectrum of the
nonunitary WZNW model which correspond to unitary highest weight states of the
Virasoro algebra \cite{Soloviev}.}.
%In what follows we will call $O$ a unitary operator.

Another important property of the operator $O$ is that it obeys the following
fusion rule
\begin{equation}
O\cdot O=[O]~+~[I]~+~..., \end {equation}
where the square brackets denote the contributions of $O$ and identity
operator $I$ and the corresponding descendants of $O$ and $I$; whereas
dots stand for operators with conformal dimensions greater than one.
The fusion given
by eq. (2.18) is easy to prove in the large $k$ limit. The proof is as follows.
The limit of large level corresponds to the classical limit of the conformal
WZNW model. In this limit, the affine-Virasoro primary field $\phi^{a\bar a}$
can be naturally identified with the classical composite field \cite{Knizhnik}
\begin{equation}
\phi^{a\bar a}=\mbox{Tr}(g^{-1}t^agt^{\bar a}).\end{equation}
So the operator $O$ goes to the classical function
\begin{equation}
O=-{k^2\over4}\mbox{Tr}(\partial g\bar\partial g^{-1}),\end{equation}
which is the sigma model term. Clearly
the operator $O$ preserves explicitly the global $G\times G$ symmetry of the
conformal WZNW model. Therefore, on the right hand side of eq. (2.18) there
must appear only primaries and their descendants which are scalars under the
given symmetry. Apparently these primaries are to be built up of the field $g$
and derivatives. Then, it is not difficult to convince oneself that there are
no
other
relevant primary operators but $O$ and $I$. Otherwise, the renormalizability of
the corresponding sigma model would be broken.

All in all, the properties of the operator $O$ are sufficient to give a unitary
renormalizable perturbation of the nonunitary conformal WZNW model. From now on
by the perturbed nonunitary WZNW model we will understand the following theory
\begin{equation}
S_{PWZNW}(\epsilon)
=S_{WZNW}(g,k)~-~\epsilon\int \mbox{d}^2z ~O(z,\bar z),\end{equation}
where $\epsilon$ is a small parameter measuring a deviation from the conformal
model $S_{WZNW}(g,k)$ \footnote{It is necessary to point out that there are two
distinct directions in which the perturbation can be performed: $\epsilon<0$
and $\epsilon>0$. In what follows we will discuss the perturbation(s) in the
(massless) direction ($\epsilon>0$).}.

We go on to compute the renormalization beta function associated to the
coupling
$\epsilon$. Away of criticality, where $\epsilon\ne0$, the beta function is
defined according to \cite{Zamolodchikov-1}-\cite{Cardy_Ludwig},\cite{Cardy}
\begin{equation}
\beta=[2-(\Delta+\bar\Delta)]\epsilon~-~\pi C~\epsilon^2~+~{\cal
O}(\epsilon^3),\end{equation}
where $(\Delta,\bar\Delta)$ are given by eq. (2.16). The constant $C$ is taken
here to be the coefficient of the three point function
\begin{equation}
\langle O(z_1,\bar z_1)O(z_2,\bar z_2)O(z_3,\bar z_3)\rangle =
C\,||O||^2\,\Pi_{i<j}^3{1\over|z_{ij}|^{2\Delta}},\end{equation}
where $||O||^2=\langle O(1)O(0)\rangle$. Let us emphasize that the two and
three point functions are calculated with respect to the $SL(2,C)$ vacuum.

One can easily solve equation (2.22) to find critical points of the beta
function. There are two solutions
\begin{equation}
\epsilon_1=0,~~~~~~~~~\epsilon_2=-{2c_V\over\pi C k}.\end{equation}
The first one obviously corresponds to the unperturbed WZNW model; whereas the
second solution signifies a new (infrared) conformal point in the vicinity of
the ultraviolet fixed point, $\epsilon_1=0$. In order to understand the meaning
of the second critical point, one has to compute the coefficient $C$ in eq.
(2.23).

Computation of coefficients of three point functions of primary fields is one
of the most difficult technical problems in conformal field theory. However, in
the case under consideration, we need only calculate it to leading order in
$1/k$. This can be done relatively easily. First of all, by using the
definition of $O$ given by eq. (2.9) we can present the three point function in
eq. (2.23) in the form
\begin{equation}
\langle O(z_1,\bar z_1)O(z_2,\bar z_2)O(z_3,\bar z_3)\rangle
=\langle\Pi_i^3\oint{\mbox{d}w_i\over2\pi i}\oint{\mbox{d}\bar w_i\over2\pi i}~
{1\over|w_i-z_i|^2}J^{a_i}(w_i)\bar J^{\bar a_i}(\bar w_i)\phi^{a_i\bar
a_i}(z_i,\bar z_i)\rangle.\end{equation}
The right hand side of the last equation can be simplified with the Ward
identities \cite{Knizhnik} to the following compact expression
\begin{equation}
\langle O(z_1,\bar z_1)O(z_2,\bar z_2)O(z_3,\bar z_3)\rangle =
\frac{k^2}{4}~f^{abc}f^{\bar a\bar b\bar c}~C^{a\bar a~b\bar b~c\bar
c}_{\phi\phi\phi}~M\Pi_{i<j}^3{1\over|z_{ij}|^{2\Delta}},\end{equation}
where $C_{\phi\phi\phi}$ is the coefficient of the three point function
\begin{equation}
\langle\phi^{a\bar a}(z_1,\bar z_1)\phi^{b\bar b}(z_2,\bar z_2)
\phi^{c\bar c}(z_3,\bar z_3)\rangle =M~C^{a\bar a~b\bar b~c\bar
c}_{\phi\phi\phi}~\Pi_{i<j}^3{1\over|z_{ij}|^{2\Delta_\phi}}.\end{equation}
The factor $M$ in eqs. (2.26), (2.27) is the matter of the $\phi$-field
normalization, $\langle\phi^{a\bar a}(1)\phi^{b\bar b}(0)\rangle
=M\delta^{ab}\delta
^{\bar a\bar b}$. It is necessary to emphasize that formula (2.26) is only
given to
leading order in $1/k$.

In order to compute the coefficient on the right hand side of eq. (2.26), one
can use the following useful observation. It turns out that in the large $|k|$
limit the operator
\begin{equation}
K^a=:\phi^{a\bar a}\bar J^{\bar a}:=\mbox{Tr}(-{k\over2}g^{-1}\bar\partial
gt^a)\end{equation}
acquires the canonical dimension (0,1). Therefore, in the quasiclassical limit
$K^a$ must behave as a current. In particular, the OPE of $K^a$ with
$\phi^{b\bar b}$ has to be as follows\footnote{From the dimension one could
argue that
\begin{eqnarray}
K^a(w,\bar w)\phi^{b\bar b}(z,\bar z)={\lambda~
f^{abc}\over\bar w-\bar z}\phi^{c\bar
b}(z,\bar z)~+~reg.\nonumber\end{eqnarray}
By comparing the last formula with the classical Poisson bracket
$\{K^a(w,t),\phi^{b\bar b}(z,t)\}$, we find $\lambda=1+{\cal O}(1/k)$.}

\begin{equation}
K^a(w,\bar w)\phi^{b\bar b}(z,\bar z)={f^{abc}\over\bar w-\bar z}\phi^{c\bar
b}(z,\bar z)~+~reg.\end{equation}
Note that one can check the last formula using the classical Poisson
brackets of the WZNW model \cite{Abdalla}. This equation gives rise to the
useful relation
\begin{equation}
C^{a\bar a~b\bar c~c\bar
d}_{\phi\phi\phi}~f^{\bar b\bar c\bar d}=f^{abc}~\delta^{\bar a\bar b}~+~{\cal
O}(1/k).\end{equation}
Taking this identity in eq. (2.26) we obtain the following expression
\begin{equation}
\langle O(z_1,\bar z_1)O(z_2,\bar z_2)O(z_3,\bar z_3)\rangle =
M~c_V~(k\dim G/2)^2~\Pi_{i<j}^3{1\over|z_{ij}|^{2\Delta}}.\end{equation}
To get the constant $C$ from the last formula, one has to normalize the two
point function to one. The norm of the operator $O$ is given by
\begin{equation}
||O||^2=\langle O(1)O(0)\rangle =M~(k\dim G/2)^2.\end{equation}
Finally we find
\begin{equation}
C={M~c_V(k\dim G/2)^2\over ||O||^2}=c_V~+~{\cal O}(1/k).\end{equation}

With the given expression for $C$ the second solution in (2.24) comes out as
follows
\begin{equation}
\epsilon_2={-2\over\pi k}.\end{equation}
It is important that the value of $\epsilon_2$ does not depend on the
normalization constant $M$.
Substituting this solution in eq. (2.21) we come to the interesting result
\begin{equation}
S_{PWZNW}(\epsilon_2)=S_{WZNW}(g,-k).\end{equation}
In other words, the WZNW model with negative level $k$ perturbed by the
operator $O$ arrives at the conformal WZNW model with positive level
$l=-k=|k|$. Thus, the two conformal points of the WZNW model discovered by
Witten \cite{Witten} turn out to be the ultraviolet and infrared fixed points
of the one renormalization group flow. It might be interesting to
understand whether there is a sort of duality symmetry between these two
conformal systems. In ref. \cite{Karabali-3} it has been observed that the
nonunitary WZNW model can come into being by duality transformations in the
space of Thirring models with positive kinetic energy.

Once we know exactly the critical theory corresponding to the perturbative
conformal point given by eq. (2.34), we can compute the exact Virasoro central
charge at the point $\epsilon_2$. We find
\begin{equation}
c(\epsilon_2)=c(\epsilon_1)~-~{2kc_V\dim G\over c_V^2-k^2}.\end{equation}
Apparently when $k<-c_V$, the difference $\Delta c=c(\epsilon_2)-c(\epsilon_1)$
is less than zero in full agreement with the $c$-theorem
\cite{Zamolodchikov-1}. We might expect this result since the perturbation was
done by the operator $O$ with positive norm.

The point to be made is that the Virasoro central charge at the infrared
conformal point $\epsilon_2$ may also be estimated by perturbation theory
(see for example, the Cardy-Ludwig formula \cite{Cardy_Ludwig}).
By comparing the perturbative result
\begin{equation}
\Delta c=-{y^3\over C^2}||O||^2=-{[2-(\Delta+\bar\Delta)]^3
||O||^2\over c_V^2}\end{equation}
with the exact expression in (2.36), we can fix the normalization constant $M$.
We find
\begin{equation}
M={1\over\dim G}.\end{equation}
The factor $||O||^2$ in eq. (2.37) comes from the Zamolodchikov metric in
the formula for the $c$-function. One can use the sigma model representation
of the operator $O$ and free field Green functions to verify the formula
for the norm.

It is instructive to compute anomalous dimensions of the operator $O$ at the
second conformal point. We can use the perturbative formula due to Redlich
\cite{Redlich}. To given order in $1/k$ the formula yields
\begin{equation}
\Delta(\epsilon_2)=\bar\Delta(\epsilon_2)=1-c_V/k\end{equation}
in full agreement with the exact result
\begin{equation}
\Delta(\epsilon_2)=\bar\Delta(\epsilon_2)=1~+~{c_V\over c_V-k},\end{equation}
where $k$ is negative.

Another observation is that the perturbation of the nonunitary
conformal WZNW model by the operator $O$ provides some insight into
the nonconformal WZNW model with arbitrary coupling constant in front of its
sigma model term. Indeed, in the large $|k|$ limit we can make use of the
quasiclassical formula (2.20) to alter the perturbed theory to the form
\begin{equation}
S_{PWZNW}(\epsilon)={1\over4\lambda}\int\mbox{d}^2x~\mbox{Tr}(\partial_\mu
g\partial^\mu g^{-1})~+~k\Gamma,\end{equation}
where $\Gamma$ is the Wess-Zumino term; whereas the sigma model coupling
$\lambda$ is related to the perturbation parameter $\epsilon$ by the formula
\begin{equation}
{1\over\lambda}={k\over4\pi}~+~{\epsilon k^2\over4}.\end{equation}
The constant $\epsilon$ runs from 0 to $\epsilon_2=-(2/\pi k)$. Correspondingly
the coupling $\lambda$ changes from $-\infty$ to $+\infty$ except the small
interval $]-(4\pi/k),~(4\pi/k)[$ which goes to zero as $|k|\to\infty$. We find
it rather amazing that inside the perturbation interval there lies the point
$\epsilon_{WZ}=-(1/\pi k)$ at which the sigma model term drops out and we are
left
with the pure Wess-Zumino term.
\begin{equation}
S_{PWZNW}(\epsilon_{WZ})=k\Gamma.\end{equation}

Some time ago classical
two dimensional models described by the pure Wess-Zumino term attracted some
attention \cite{Park-2}. Now we have exhibited that such peculiar theories can
be
properly understood at the quantum level as the massive perturbation of the
nonunitary WZNW model. Our conjecture is that the pure Wess-Zumino term theory
separates the nonunitary phase from the unitary phase of the WZNW model
given by eq. (2.41). That is, at this point, all states with negative norms
have to turn into null vectors, perhaps due to appearance of the additional
symmetry. We are going to explore this point elsewhere.

\section{The sigma model perturbation of the gauged WZNW model}

In this section we would like to consider one application of the
sigma model perturbation of the nonunitary WZNW model. It is the GWZNW model
in which the nonunitary WZNW theory emerges as an intrinsic component (see e.g.
ref.\cite{Karabali_Schnitzer}). Indeed, at the classical level the GWZNW model
can be described as a combination of usual conformal WZNW models
\begin{equation}
S_{GWZNW}=S_{WZNW}(hg\tilde h;k)~+~S_{WZNW}(h\tilde h;-k).\end{equation}
Here the matrix field $g$ takes its values on the Lie group $G$; whereas
$h,~\tilde h$ take values on the subgroup $H$ of $G$. Apparently one of these
two WZNW models has to be nonunitary. The usual action of the GWZNW model in
terms of the group element $g$ and the vector nonabelian fields $A_z$ and $\bar
A_{\bar z}$ is obtained from eq. (3.44) upon using the following definition
\begin{equation}
A_z=h^{-1}\partial h,~~~~~~~~\bar A_{\bar z}=\bar\partial \tilde h\tilde
h^{-1}.\end{equation}
Correspondingly the gauge symmetry in the variables $g,~h,~\tilde h$ is given
by
\begin{equation}
g\to\Omega g\Omega^{-1},~~~~~h\to h\Omega^{-1},~~~~~\tilde h\to\Omega\tilde
h,\end{equation}
where $\Omega$ is the parameter of the gauge group $H$.

At the quantum level, the GWZNW model is described by the action
\begin{equation}
S_{QGWZNW}=S_{WZNW}(hg\tilde h;k)~+~S_{WZNW}(h\tilde
h;-k-2c_V(H))~+~S_{Gh}(b,c,\bar b,\bar c).\end{equation}
Compared to the classical expression, the quantum action has the second
(nonunitary) WZNW model of the product $h\tilde h$ with the level shifted by
twice the eigenvalue of the quadratic Casimir operator in the adjoint
representation of the subalgebra ${\cal H}$. In addition, the QGWZNW model has
the ghost-like contribution
\begin{equation}
S_{Gh}=\mbox{Tr}\int\mbox{d}^2z~(b\bar\partial c+\bar b\partial\bar
c).\end{equation}
The modifications to the quantum action have a very elegant explanation
\cite{Hwang}. All these changes in the quantum theory work to convert the
second class constraints of the classical GWZNW model into first class
constraints. A systematic investigation of the conversion was first discussed
in
\cite{Batalin} and then extensively studied in \cite{Batalin_Tyutin}.
The theory also has a nilpotent BRST operator \cite{Karabali_Schnitzer}.

The point to be made is that the product $h\tilde h$ is a gauge invariant
quantity. So is any function of $h\tilde h$. Moreover, the gauge symmetry
allows us to impose the following gauge condition
\begin{equation}
\tilde h=1,\end{equation}
which does not lead to any additional propagating Faddeev-Popov ghosts.
Therefore, any function of $h$ has to respect the gauge symmetry. Let us
consider the following one
\begin{equation}
O_H=:J^a\bar J^{\bar a}\phi^{a\bar a}:,\end{equation}
where all the three operators on the right hand side are defined in terms of
the nonunitary level $(-k-2c_V(H))$ WZNW model on $H$:
\begin{eqnarray}
\phi^{a\bar a}&=&\mbox{Tr}(h^{-1}t^aht^{\bar a}),\nonumber\\
J^a&=&{1\over2}\eta^{ab}\mbox{Tr}[(k+2c_V(H))h^{-1}\partial ht^a],\\
\bar J^a&=&{1\over2}\eta^{ab}\mbox{Tr}[(k+2c_V(H))\bar\partial
hh^{-1}t^a].\nonumber\end{eqnarray}
Normal ordering is understood in accordance with eqs. (2.9), (2.10).
The operator $\int d^2z~O_H(z,\bar z)$
respects the gauge symmetry (3.46) in the sense that being
substituted into any correlation functions it will not spoil the Ward identity
attributed to the gauge symmetry given by (3.46). Namely,
\begin{equation}
\langle\partial\bar J^{tot}(\bar z)~...~\int d^2w~O_H(w,\bar w)~...
{}~\rangle~-~\langle\bar\partial
J^{tot}(z)~...~\int d^2w~O_H(w,\bar w)~...~\rangle=0,\end{equation}
where
\begin{eqnarray}
J^{tot,\,a}={1\over2}\mbox{Tr}[(k+2c_V(H))h^{-1}\partial ht^a]~-~
{1\over2}\mbox{Tr}(kg^{-1}\partial gt^a)~-~f^{abc}b_z^bc^c
\nonumber\\ & & \\
\bar J^{tot,\,a}={1\over2}\mbox{Tr}[(k+2c_V(H))\bar\partial
hh^{-1}t^a]~-~{1\over2}\mbox{Tr}(k\bar\partial gg^{-1}t^a)~-~
f^{abc}\bar b_{\bar z}^b\bar c^c.\nonumber\end{eqnarray}
We have checked explicitly \cite{Soloviev-2} the Ward identity (3.52)
to leading order in $1/k$.

The given operator $O_H$ is a good physical operator as it gives rise to a
unitary Virasoro representation in the GWZNW model. This operator
possesses all the merits of the operator $O$ considered in the previous
section. Therefore, we can make use of $O_H$ to perturb the GWZNW model.
Henceforth we will call perturbed GWZNW model the following theory
\begin{equation}
S_{PGWZNW}(\epsilon)=S_{QGWZNW}~-~\epsilon\int \mbox{d}^2z~O_H(z,\bar z).
\end{equation}

Due to the striking analogy between $O_H$ and $O$ from the previous
section, we can conclude that the PGWZNW model given by eq. (3.54) has to have
a second
(infrared) conformal point. The value of the parameter $\epsilon$ at this point
can be deduced from eq. (2.34). To leading order in $1/k$ we find
\begin{equation}
\epsilon_2={2\over\pi k}.\end{equation}
The quantity $\epsilon_2$ occurs with a positive sign because $k$ - the level
of the ungauged
WZNW model - is a positive integer in the case under consideration.

The exact conformal theory corresponding to the perturbative fixed point
can be recognized as follows
\begin{equation}
S_{PGWZNW}(\epsilon_2)=S_{WZNW}(g;k)~+~S_{WZNW}(h;k+2c_V(H))~+~S_{Gh}.
\end{equation}
Correspondingly the Virasoro central charge at the infrared critical point is
given by
\begin{eqnarray}
c(\epsilon_2)&=&c(G/H)~+~c_{WZNW}(k+2c_V(H))~-~c_{WZNW}(-k-2c_V(H))\nonumber\\
& & \\
&=&c(G/H)~-~
{2(k+2c_V(H))c_V(H)\dim H\over k^2+4kc_V(H)+3c_V(H)^2}.\nonumber\end{eqnarray}
Here $c(G/H)$ is the Virasoro central charge of the $G/H$ coset construction;
whereas $c_{WZNW}(l)$ denotes the Virasoro central charge of the level $l$ WZNW
model on the subgroup $H$. It is clear from eq. (3.57) that
$c(\epsilon_2)<c(G/H)$ as it should be according to Zamolodchikov's
$c$-theorem.

Both relation (3.56) and (3.57) indicate that the second (infrared) critical
point no longer corresponds to a gauged WZNW theory. It is not surprising
because the perturbation by the operator $O_H$ breaks the BRST invariance of
the GWZNW model. This may have some interesting consequences. First of all, we
want to point out that all ghost free states of the GWZNW model after
perturbation by the operator $O_H$ have to go to unitary states at the
infrared conformal point. Indeed, it follows from eq. (3.56) that
at the perturbative conformal point
any ghost free state belongs to the direct product of
Fock spaces of the unitary WZNW models. Since the perturbing operator does not
involve the ghosts, all ghost free states of the GWZNW model go to ghost free
states of the perturbed theory. It was observed in ref. \cite{Karabali}
that ghost free states of the GWZNW model may be representatives of the
cohomology classes forming the physical subspace $\mbox{Ker}Q/\mbox{Im}Q$ (with
$Q$ the BRST operator) of the GWZNW model. If it is the case (at least for some
particular models), then it may be
very interesting to investigate what is happening to the fusion algebra of
these ghost free representatives
under the perturbation described by eq. (3.54). Apparently such unitary primary
states
have to continue being unitary vectors at the perturbative conformal point.
By unitary states we mean states which have positive definite inner product
with themselves.
Of particular interest are GWZNW models with a finite number of cohomology
classes such as $SU_k(2)\times SU_1(2)/SU_{k+1}(2)$ models. It might be
important to understand the mechanism of deformation of fusions of these models
in the course
of the $O_H$ perturbation and to see how (an infinite number of) new primaries
will enter deformed fusion rules at the infrared conformal point.

\section{The BRST invariant sigma model perturbation of the GWZNW model}

We have described in the previous section the operator $O_H$ which is suitable
for performing relevant unitary deformations of GWZNW models. However, this
operator $O_H$ as it was defined in eq. (3.50) has no proper action on the
physical subspace
$\mbox{Ker}\,Q/\mbox{Im}\,Q$ of the GWZNW model. Yet it might be interesting to
have perturbations genuinely defined on the space $\mbox{Ker}\,Q/\mbox{Im}\,Q$.
The aim of the present section is to exhibit an operator which provides
relevant perturbations on the given space.

Let us start with the nilpotent BRST operator $Q$ of the given GWZNW model.
The nilpotent BRST operator $Q$ is defined as follows
\cite{Karabali_Schnitzer}
\begin{equation}
Q=\oint{\mbox{d}z\over2\pi i}[:c^a\,(\tilde
J^a+J^a):(z)~-~{1\over2}f^{abc}:c^a\,b^b\,c^c:(z)],\end{equation}
where we have used the following notations
\begin{eqnarray}
J&=&{(k+2c_V(H))\over2}h^{-1}\partial h,\nonumber \\ & & \\
\tilde J&=&-{k\over2}g^{-1}\partial g|_H.\nonumber\end{eqnarray}
Here the current $\tilde J$ is a projection of the ${\cal G}$-valued current on
the subalgebra ${\cal H}$ of ${\cal G}$. Note that our notations are slightly
different from those in \cite{Karabali_Schnitzer}.

It is instructive to verify that the operator $O_H$ in eq. (3.50) is not
annihilated by $Q$. However it would be the case if the current $J$ in the
definition of $O_H$ had vanishing central charge.
The last observation gives a hint at the way one has to alter
$O_H$ to end up with a BRST invariant operator.

Another crucial point is that the currents $J$ and $\tilde J$ form the affine
algebras with central elements of the opposite signs. As a matter of fact,
there must exist a linear combination of $J$ and $\tilde J$ which acting on
$\phi$ has to give rise to an
affine primary vector with respect to the affine current $J+\tilde J$ from $Q$.
Indeed, one can check that the following relations
\begin{equation}
(J^a_{n\ge0}+\tilde J^a_{n\ge0})\cdot (J^b_{-1}+{(k+4c_V(H))\over k}
\tilde J^b_{-1})\phi^b|0\rangle=0
\end{equation}
hold. Thus, $(J^b_{-1}+{(k+4c_V(H))\over k}
\tilde J^b_{-1})\phi^b|0\rangle$ is an affine primary scalar.

After that, it is easy to prove that the following operator
\begin{equation}
O_{BRST}=:[J^a+{(k+4c_V(H))\over k}\tilde J^a]
[\bar J^{\bar a}+{(k+4c_V(H))\over k}\tilde{\bar J^{\bar a}}]\phi^{a\bar
a}:,\end{equation}
where $\phi^{a\bar a}$ is as in eq. (3.51), is BRST invariant. That is,
\begin{equation}
[Q,O_{BRST}(z,\bar z)]=\oint{\mbox{d}w\over2\pi i}~:c^a(\tilde
J^a+J^a):(w)~O_{BRST}(z,\bar z)=0,\end{equation}
where the contour surrounds the point $z$.
%Note that the currents $\tilde J$
%and $\tilde{\bar J}$ enter eq. (4.61) with the coefficient $(k+c_V(H))/k$, not
%with $(k+2c_V(H))/k$. The difference in the coefficients is because of
%normal ordering in eq. (4.61).
Obviously
$O_{BRST}$ has essentially
the same algebraic structure as $O$ and $O_H$.

The point to be made is that in the classical limit $|k|\to\infty$ the
operator $O_{BRST}$ can be presented in the form
\begin{eqnarray}
O_{BRST}=:J'^a~\bar J'^{\bar a}:
,\nonumber\end{eqnarray}
where
\begin{eqnarray}
J'&=&-{k\over 2}(gh^{-1})^{-1}\partial(gh^{-1}),\nonumber \\
\bar J'&=&\tilde {\bar J}~+~\bar J
=-{k\over2}(\bar\partial gg^{-1}|_H-\bar\partial hh^{-1}).\nonumber
\end{eqnarray}
The last formulas might be useful in the course of Lagrangian interpretation of
the operator $O_{BRST}$. It shows that $O_{BRST}$ to certain extent can be
understood as a Thirring-like interaction between the currents $J'$ and
$\bar J'$. It is necessary to point out that in the classical theory, the
current $\bar J'$ coincides with the classical gauge constraint. So that at the
classical level, $O_{BRST}$ goes to zero.

Also it is obvious that $O_{BRST}$ and $O_H$ share the same
conformal dimensions, so that in the large $k$ limit, the given operator
$O_{BRST}$ has to behave as a relevant quasimarginal operator. It still
provides a
unitary conformal representation
since $O_{BRST}$ does not contain the ghost-like fields. Furthermore,
$O_{BRST}$ has to continue satisfying the fusion algebra
\begin{equation}
O_{BRST}\cdot O_{BRST}=[O_{BRST}]~+~[I]~+~...,\end{equation}
where the dots represents contributions of operators with irrelevant conformal
dimensions (lager than one).

Indeed, in order to prove eq. (4.63) in the large $k$ limit, we can again turn
to the symmetry and renormalizability arguments as
we did it in the case of the operator $O$ as well as $O_H$.

All in all, the operator $O_{BRST}$ appears to be an appropriate relevant
quasimarginal unitary operator for carrying out perturbations from the physical
subspace $\mbox{Ker}\,Q/\mbox{Im}\,Q$ of the GWZNW model.

We go on to define the BRST invariant perturbed GWZNW model
\begin{equation}
S_{BRST}(\epsilon)=S_{QGWZNW}~-~\epsilon~\int\mbox{d}^2z~O_{BRST}.
\end{equation}
We want to point out that in accordance with eq. (4.61), the given perturbation
around the GWZNW model can be
understood as a linear combination of two perturbations by the sigma model term
and by a Thirring-like current-current interaction. The latter
introduces novel
features in computations with $O_{BRST}$. We can no longer rely on
results obtained for the operators $O$ and $O_H$ from the previous sections.
This is mainly because of the fact that the linear combination
$J+{(k+4c_V(H))\over k}\tilde J$ in the large $k$ limit behaves as an affine
current with central charge $3c_V(H)$. Therefore, the central term
of the underlying affine algebra no longer dictates the
leading order in $1/k$ and,
hence,
more intricate computations are required. Most of the difficulties reside in
the
three point function of the spin (1,1) operator $\phi^{a\bar a}$ from the
nonunitary WZNW model. To date,
three point functions of all affine primaries have been computed only in the
case of the $SU(2)$ WZNW model \cite{Dotsenko}.
Fortunately the identity (2.30) allows us to overcome most of the technical
obstacles.
%All the above mentioned technical
%obstacles drive us to consideration of special $G/SU(2)$ GWZNW models. The
%latter are actually good examples to study the effects of our perturbation.

Let us consider one very interesting GWZNW model with $G=SU_{m-2}(2)\times
SU_1(2)$ and $H=SU_{m-1}(2)$. The physical subspace
$\mbox{Ker}\,Q/\mbox{Im}\,Q$
of this theory properly describes the Fock space of the minimal conformal model
with the Virasoro central charge \cite{Hwang}
\begin{equation}
c=1~-~{6\over m(m+1)},~~~~~~m=3,4,...\end{equation}

It is straightforward to compute the conformal dimensions of the corresponding
operator $O_{BRST}$. We find
\begin{equation}
\Delta=\bar\Delta=1-{c_V(SU(2))\over m-1+c_V(SU(2))}=1-{2\over m+1}.
\end{equation}
There is one primary operator from $\mbox{Ker}\,Q/\mbox{Im}\,Q$ with the given
conformal dimensions. This is the $\Phi_{(3,1)}$ operator. Since $O_{BRST}$
belongs to the same physical subspace, we have to identify $O_{BRST}$ with
$\Phi_{(3,1)}$ (up to $Q$-exact terms). Note that the fusion algebra of
$\Phi_{(3,1)}$ agrees with eq. (4.63). This way we learn that our perturbation
of the GWZNW model coincides with the  perturbation of the
minimal conformal series first discovered by Zamolodchikov
\cite{Zamolodchikov-1},\cite{Cardy}. It is rather amazing that the
$\Phi_{(3,1)}$ perturbation, in fact, stems from the relevant perturbation
around the nonunitary WZNW model \cite{Soloviev-1}. With this in mind, formula
(4.61) exhibits a microscopical structure of the $\Phi_{(3,1)}$ operator.
Surprisingly the value of the perturbation parameter $\epsilon$
at the infrared conformal point in the theory in eq. (4.64) is still given by
formula (3.55). On the other hand for
the perturbative Virasoro central charge formula
(2.37) gives rise to the expression
\begin{equation}
\Delta c=-{2c_V(H)\dim^2H\over m^3}~M,\end{equation}
where we have used the following result
\begin{equation}
||O_{BRST}||^2= M~\dim^2 H/4.\end{equation}
Here the normalization constant $M$ is given by
eq. (2.38). In order to derive the last formula, one has to
make use of the definition of $O_{BRST}$ in eq. (4.61) with normal ordering as
in eq. (2.9).
In the case under consideration, one finds
\begin{equation}
\Delta c=-{12\over m^3}~+~{\cal O}(m^{-4}).\end{equation}
The last formula agrees with the exact result for the given coset. It is
interesting that the normalization constant $M$ is universal for all cosets and
is fixed in the nonunitary WZNW model.
Apparently, the infrared conformal point obtained by
perturbation of the minimal conformal
model, guarantees the existence of the second conformal point at the same value
of $\epsilon$ for all $G/SU(2)$ GWZNW models with arbitrary $G$. By computing
the Virasoro central charge at the infrared conformal point, we establish the
possibility of flow between the following coset constructions
\begin{equation}
{\tilde G_1\times SU_{k-1}(2)\over SU_k(2)}\to{\tilde G_1\times
SU_{k-2}(2)\over SU_{k-1}(2)}.\end{equation}

%In order to exhibit flows between arbitrary $G/H$ cosets, we have to overcome
%some technical problems. We hope to accomplish such computations, at least,
%%for
%some GWZNW theories which are distinct from $G/SU(2)$ models. In particular,
%%it
%is interesting to
By using our perturbation on the GWZNW model, we can easily
prove the conjecture about the flow between symmetric
$G_k\times G_l/G_{k+l}$ and $G_{k-l}\times G_l/G_k$ cosets (see e.g. \cite{Ahn}
).  Indeed, for arbitrary $G$ we find (in the large $k$ limit)
\begin{equation}
\Delta c=-{2c_V(G)\dim G~l^2\over k^3}~+~{\cal O}(1/k^4).\end{equation}
One can verify that the given difference agrees with the exact result for the
coset constructions under consideration.
Note that our BRST invariant perturbation coincides with the perturbation
operator of the $G_k\times G_l/G_{k+l}$ cosets considered in \cite{Ahn}.
Therefore, it is suggestive that $S$-matrices for perturbed gauged WZNW models
can be related to the restricted ${\cal R}$-matrices of the Toda field
theories \cite{Ahn}.

%For general $G/H$ coset,
%the BRST invariant perturbation gives rise to the difference
%bewteen the Virasoro central charges
%\begin{equation}
%\Delta c=-{2c^3_V(H)\dim H\over k^3}~+~{\cal O}(1/k^4)\end{equation}
%which corresponds to the following flow
%\begin{equation}
%{G_k\over H_k}\to{G_k\times H_{k+c_V(H)}\times H_{k-c_V(H)}\over H_k\times H_k
%\times H_k}.
%\end{equation}
Because coset constructions describe certain geometries it makes sense to
introduce a notion of geometry flow along the trajectory of the renormalization
group.

\section{Conclusion}

We have started with the relevant perturbation around the nonunitary WZNW model
and proceeded to define proper deformations of gauged WZNW models as well as
general coset constructions.
We found that our perturbations being generated by the operators agree
with the Zamolodchikov $c$-theorem. We have established that the BRST invariant
perturbing operator in the case of the $SU_k(2)\times SU_1(2)/SU_{k+1}(2)$
GWZNW model
coincides with the $\Phi_{(3,1)}$ operator of the minimal conformal series. At
the same time, our perturbation operator seems to be equally suitable for
performing relevant hermitian perturbations around general GWZNW models. In
this way one can discover
renormalization group flows between different conformal theories describing
different geometries of the target space.

\par \noindent
{\em Acknowledgement}: I would like to thank J. M. Figueroa-O'Farrill,
M. Green, C. M. Hull, E. Ramos, A. Semikhatov, I. Vaysburd and G. M. T. Watts
for useful
discussions. I am indebted to the referee for helpful comments.
I thank also J. U. H.
Petersen and S. Thomas for reading the manuscript.
I would also like to thank the SERC and the European Commision Human Capital
and Mobility Programme for financial support.

\end{document}